# Laser-based fabrication of microstructures on nickel thin films

Srikanth Itapu[*1], Vamsi Borra[2], and Daniel G. Georgiev[1]

[1]Department of Electrical Engineering and Computer Science, University of Toledo, Toledo, Ohio, U.S.A. 43606

[2] Department of Applied Engineering and Technology, California University of Pennsylvania, California, PA, U.S.A 15419

[*]Corresponding E-mail: srikanth.itapu@rockets.utoledo.edu

**Abstract**

This work reports on the fabrication of microbump structures on Ni films by single-pulse, localized laser irradiation. Conditions for the reproducible formation of such microstructures have been identified in terms of laser-irradiation and film parameters after systematic studies involving a relevant parameter space. The cracks and voids morphology of the sputtered films was rendered undesirable and hence, smoother Ni thin film of same thickness (200nm) were deposited by vacuum evaporation. The continuous nature of the film resulted in radially symmetric thermal expansion and deformation, thus achieving a high yield of microstructures. An improvement in the inductance and the quality factor of on-chip spiral inductors incorporating such laser-microstructured ferromagnetic nickel thin films was observed, which demonstrates the potential of such a laser-based method for fabrication or fine tuning of various micro-/nano-electric/electronic sensor and other components and systems.

## 1. INTRODUCTION

In the wake of rapid growth in wireless communications, low-cost and high performance compact integrated circuit components such as on-chip inductors play a major role in circuit performance [1]. On-chip inductors often consume large area for moderate inductance (L) values and have relatively low-quality factor (Q). Besides reducing the physical circuitry of on-chip inductors, enhanced L and Q are also required in radio-frequency (RF) applications. Various approaches to overcome such limitations have been addressed in recent years, such as incorporating magnetic materials [2][3][4][5][6][7][8], laminating and patterning ferromagnetic thin films [9][10][11][12][13], utilizing in-plane and out-of-plane anisotropy to enhance magnetic fields [14], patterning ground shields [15][16], fabricating multi-layers of magnetic thin films [17][18][19], etc. In this paper, we report on the possibility of forming microbump structures on films of magnetic metals, such as Ni, by single-pulse localized laser irradiation. We identified laser, geometry, and film quality conditions under which fabrication of such microstructures is possible and then examined this technique as a method to improve/enhance the L and Q of on-chip spiral inductors. The pulsed-laser irradiation technique offers the advantage of localized thermal heating, noncontact nature and high throughput as compared to conventional microstructuring methods. Microstructuring on various metal films have been studied and different theoretical models have been proposed in recent years [20]–[22].

Metrics that primarily define the performance of magnetic materials and related devices are lower core loss, higher saturation flux density, higher anisotropy, lower magnetostriction and higher permeability. Many ferromagnetic materials (alloys, composites) along with metals (i.e., Ni, Fe, Co) have been under investigation in order to obtain an optimized L and Q performance. Our emphasis is to utilize the capabilities of laser irradiation to form microstructures on a basic ferromagnetic metal, Ni, which could then be expanded into other ferromagnetic metals and alloys to obtain further enhancement of device structures.

In addition to choosing an appropriate magnetic material, it is important to demonstrate the industrial relevance of depositing and using such materials. Commonly used techniques deposition techniques include sputtering, electro-deposition and vacuum evaporation. Sputtering has the advantage of depositing a wide range of magnetic alloys and oxide thin films of thickness that can be from a fraction of a micrometer to several micrometers and of reasonable resistivity. Similarly, vacuum evaporation and electro-deposition are commonly used techniques, with the latter preferred for depositing thicker films. Vacuum evaporation is relatively inexpensive for depositing commonly used soft magnetic materials, compatible with standard IC fabrication [23]. In this paper, we deposited Ni thin films that are a few hundred nanometers thick encapsulating a Cu spiral inductor deposited and patterned on a glass substrate. The morphology of the Ni films, which, in general, depends on the deposition technique and conditions, affects the processes that occur in the film upon high-intensity laser irradiation. Therefore, laser microstructuring of Ni thin film deposited by different techniques can be controllable to different degrees, which is why two types of Ni films were used in this study: deposited by either sputtering or vacuum evaporation.

The fabrication of microstructures on other, non-magnetic metals by nanosecond pulsed-laser irradiation using a Nd:YAG Q-switched laser, operated at its fourth harmonic wavelength ($\lambda$ =266nm) was reported relatively recently and was found to be an efficient tool as compared to alternative techniques[24], [25]. The increase in L and Q due to microstructuring was predicted by simulating a layered structure of on-chip spiral inductors [20].

## 2. EXPERIMENTAL
*2.1 Substrate preparation*

In order to obtain laser irradiated microstructures with high yield, a thermally insulating substrate is desirable which helps confine the heat flow laterally, within the metal film. Hence, microscope soda-lime glass slides were used as a substrate material. Our general substrate cleaning procedure is as follows: wash in cleaning solution (Micro-90), then rinse with DI water, followed by ultra-sonication bath in methanol for 20-25 min, and, finally, ultra-sonication bath in ethanol for 20-25 min. In between these steps, the surfaces are rubbed with lint-free wipe and blown dry with nitrogen.

*2.2 Ni deposition*

Ni thin films were prepared by both vacuum evaporation and RF magnetron sputtering. However, we have noticed that the Ni films obtained from the evaporation resulted in much smoother surfaces when compared to the sputtered films (see Fig 1). We also observed that the smooth surfaces from the evaporation aided in the Ni bump formation.

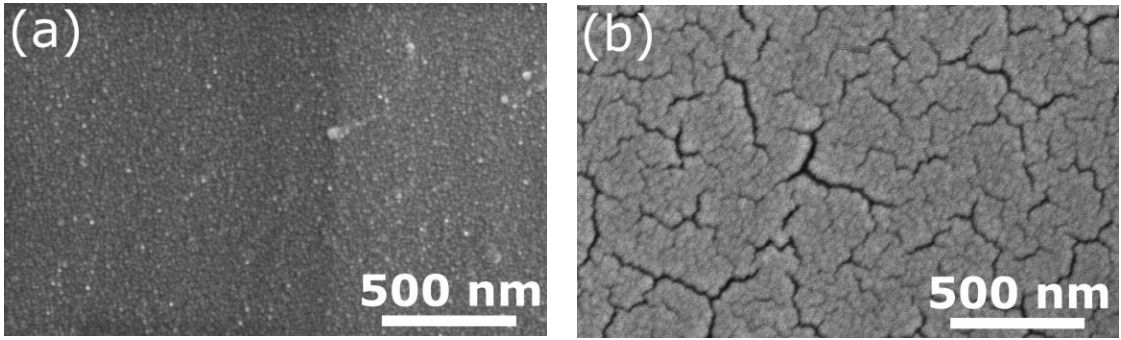

**Figure 1.** Ni films obtained by (a) evaporation (b) sputtering.

**Evaporation**: 99.98% pure Ni pellets (from Kurt J. Lesker) are evaporated from alumina coated tungsten boats (EVS9AAOW from Kurt J. Lesker) onto the substrates. Evaporating Ni using regular tungsten boat is very ineffective and unpredictable. Due to the relatively high melting point (of 1453°C), the Ni alloys with the tungsten boat as soon as it melts, causing boat to become delicate and break. In the alumina coated boat, alumina serves as a barrier between the Ni and tungsten, thus extending the life of the boat. Nevertheless, we found that the alumina coated boats lasted for 3 – 4 runs before they start to break.

The evaporation was carried out in a Denton Vacuum machine (model DV-502A) at base vacuum of $(1.6 – 2.5) \times 10^{-6}$ Torr. Before the actual evaporation, a vacuum chamber cleaning procedure is performed by gradually increasing the current through an empty tungsten boat to a high value of 250 A and keeping it on for at least 10 min. Followed by a cooling step, the evaporator chamber is brought to atmospheric pressure by purging it with dry nitrogen. Ni pellets are then placed in the alumina coated tungsten boat along with the glass substrate for the actual evaporation. Once the pressure in the vacuum chamber reaches the $10^{-6}$ Pa range, the source shutter is kept closed while the current is gradually increased until the evaporation point is reached. The metal is let to boil for 30 – 45 sec before opening the shutter. 200nm thick films are deposited at a deposition rate of 1 – 3 Å/sec.

Radio-Frequency (RF) sputtering was carried out using a 99.99% pure 3"Ni target (Kurt J. Lesker) in a Torr International sputtering system with a base vacuum of 8.0 x $10^{-7}$ Torr and sputtering pressure of 15mTorr, which, at a power level of 100W resulted in a deposition rate of 0.5 Å/sec , again to a final thickness of 200nm.

*2.3 Film Characterization Methods*

The surface morphology of the thin film samples was examined by using scanning electron microscopy (SEM) in a Hitachi S-4800 machine, operated in secondary-electron mode with acceleration voltage of 5 kV in order to limit the observation only to the film. The energy dispersive X-ray spectrograph (EDS) results confirmed that the composition of the metal layers matched the composition of the sputtering targets or the evaporation pellets. The thickness of the films was confirmed by step height measurements using a stylus profilometer (Veeco Dektak 150).

*2.4 Measurement of L and Q*

A vector network analyzer (VNA) is used to characterize any two-port network by measuring its S-parameters. From [20], The L and Q values were obtained by converting S-parameters into Y-parameters and a simulated comparison was established for the values obtained with and without modeling microstructures on a ferromagnetic thin film. In this paper, we obtain L and Q values for two samples of fabricated on-chip spiral inductors: one without laser microstructuring and the other with laser microstructuring with the help of the VNA. A standard 50Ω small miniature aperture (SMA) connector is used to evaluate two-port S-parameters by means of a HP 8720B network analyzer operating with a frequency range of 130 MHz to 20 GHz. The standard Short-open-load-thru (SOLT) calibration is performed for measuring the estimated Q [26].

## 3. RESULTS AND DISCUSSION

*3.1 Fabrication of Ni microbumps on Spiral Inductor stack*

The spiral inductor stack is fabricated as shown in Fig. 2 in the following sequence: First, a 1μm thick copper film is deposited on a soda-lime glass substrate over which a 400nm $SiO_2$ film was deposited by RF sputtering. A nickel film of thickness 200nm was then deposited by RF sputtering on one type of samples and by vacuum evaporation on the other. Laser-assisted microstructuring is performed by a step-and-repeat projection mask process resulting in a 8 x 150 array of Ni microstructures, each having a spot diameter of 10μm and vertical height of around 4μm on a sample area of 0.8cm by 0.8cm [21][22].

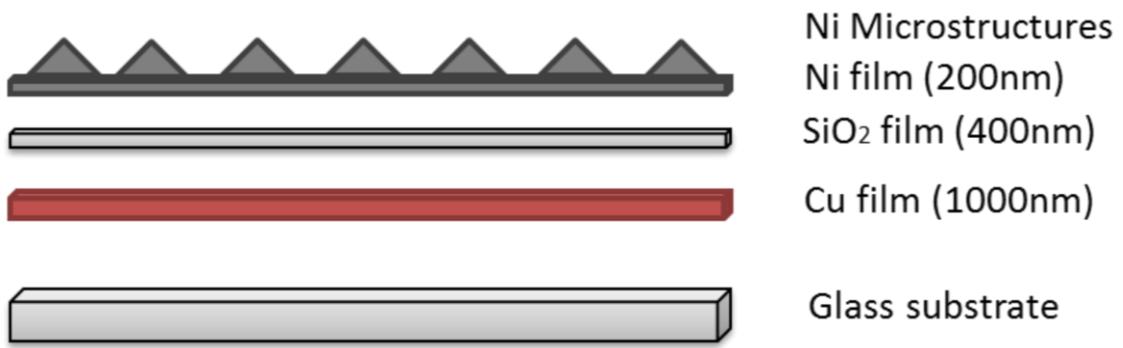

**(a)**

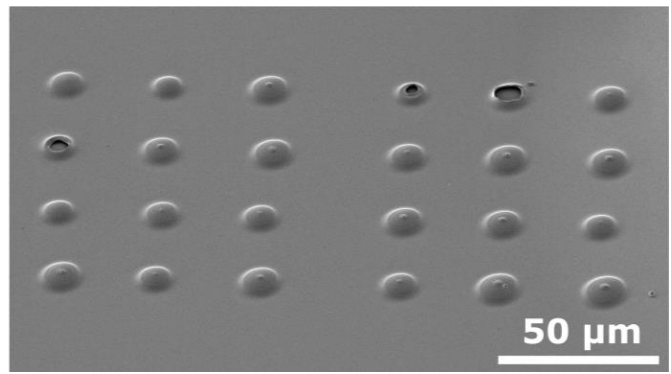

**(b)**

**Figure 2(a).** Layered view of the spiral on-chip inductor, **(b)** Microstructured Ni film on top of the spiral inductor.

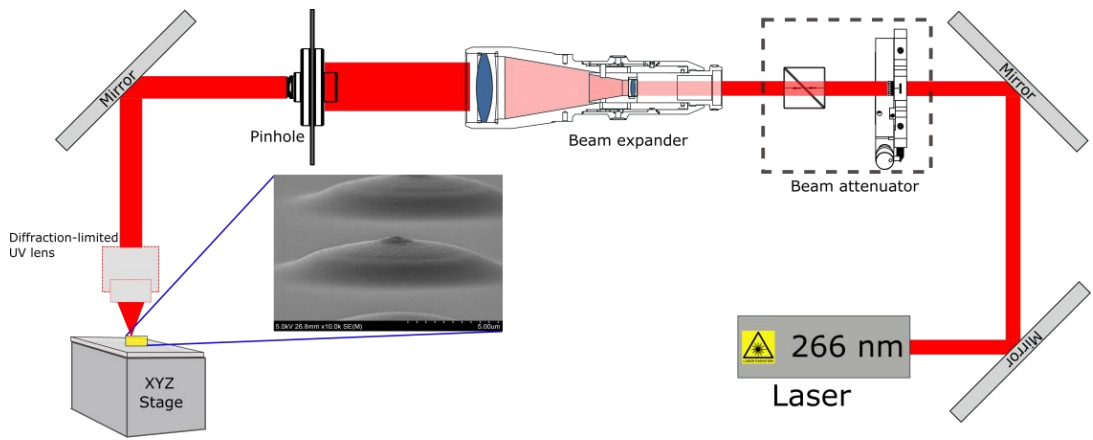

**Figure 3.** Experimental setup for the laser fabrication of Ni microbumps.

The experimental setup for laser irradiation on the spiral inductor stack is as shown in Fig.3. The 4$^{th}$ harmonic wavelength ($\lambda = 266$nm) of the Nd:YAG laser was used as the operating wavelength for formation of microbumps and it could provide nanosecond pulses with laser fluence of up to 1

J/cm$^2$. Pin holes with diameters of 25μm, 50μm, 75μm, or 100μm were used as masks along with a diffraction-limited UV objective lens, operated at a demagnifying ratio of 10:1. The beam attenuator is used to tune the laser fluence that can be sufficient for material melting. The beam expander is used to obtain a linear profile of the Gaussian beam emitting from the Nd:YAG laser. Thus, we obtained irradiation spots and the resulting microbumps with diameters of about 5μm, 7.5μm, or larger upon optimizing the laser energies corresponding to the size of the mask used. The XYZ stage provides for proper focusing/imaging, lateral positioning, and the step-and-repeat sequence.

*3.2 Ni microbumps obtained on RF sputtered Ni thin film*

SEM images (Fig. 4 and 5) of Ni thin film deposited by *RF sputtering* confirm circular laser spots with spot size in agreement with the estimated 2.5μm (Fig. 4) and 5μm (Fig. 5) base diameter and height of microbumps of about 2μm. Also, a grain structure with a fine network of inter-grain gaps (or cracks) can be seen from the SEM images. This structure is typical of all sputtered Ni films that we were able to obtain regardless of the specific conditions of deposition, which is reported by other researchers as well [27], [28].

The formation of microbumps and similar structures that develop on metal films on substrates that are less thermally conducting as a result of micrometer-scale localized pulsed laser irradiation has been discussed elsewhere[21], [24], [25], [29], [30], [31]. Similar scenarios, related to how the laser-deposited heat within the irradiated spot propagates laterally and vertically (and what changes to the film material would result), can be used to interpret the results from Ni films irradiation in this work. The inter-grain voids/cracks that exist in the sputtered films would lead to relatively poor heat conduction in a lateral (within the plane of the film) direction, and there is limited heat conduction vertically as well (air on top and thermally insulating substrate material below). As a result, a significant amount of heat will be trapped within the laser irradiated spot causing rapid thermal expansion (probably a shock wave as well) that fractures the film, helped by the existing cracks (see Fig. 4(a), 5(a) and 5(b)). At even higher laser pulse energies the center of the spot, which is the hottest part, undergoes partial melting (local spots melt within the central region) and ablation as can be seen from Fig. 4 (b-d), 5(c) and 5(d). Because of the large exposed grain surface area, partial oxidation of the central part of the spot is likely accompanying the ablation and the melting processes.

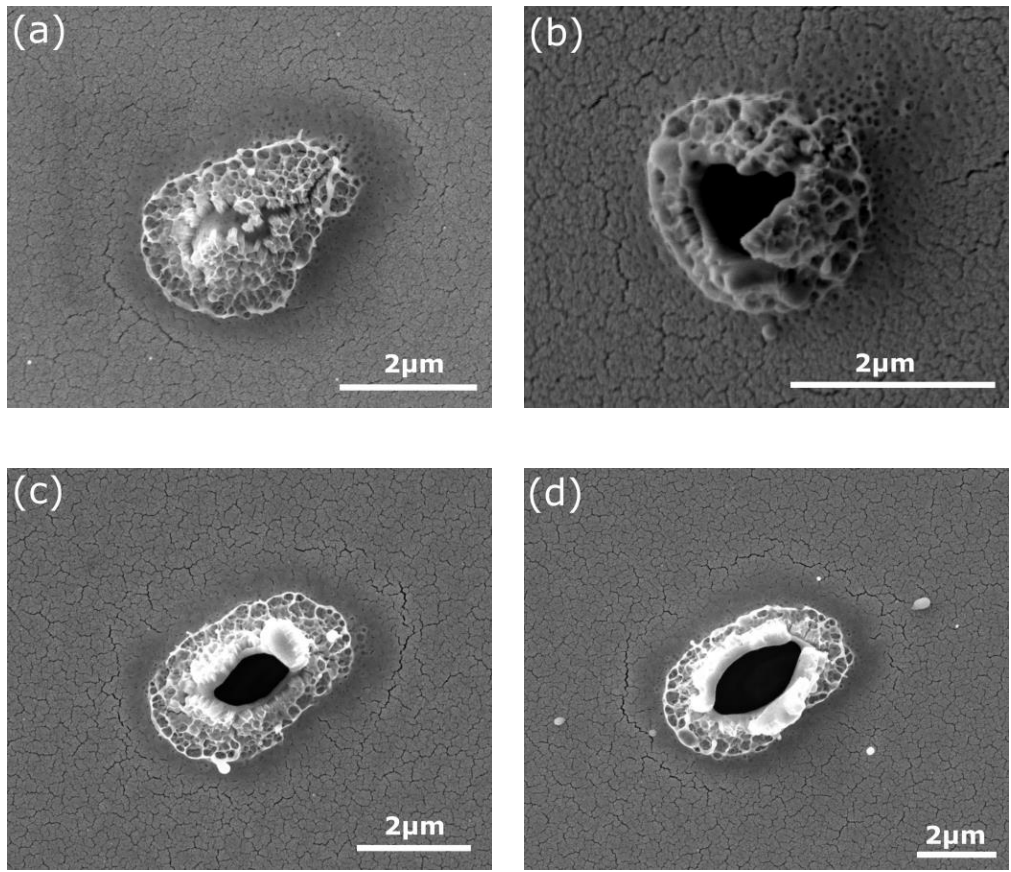

**Figure 4.** SEM images illustrating microbump formation on sputtered Ni films of thickness 200nm as a function of the energy density of the laser pulse for an irradiation spot size of 2.5µm: **(a)** 0.4 J/cm$^2$, **(b)** 0.45 J/cm$^2$, **(c)** 0.5 J/cm$^2$ and **(d)** 0.55 J/cm$^2$ single pulse.

Microbumps formation associated with molten material accumulation caused by the surface tension gradient can be observed in fig. 4(a). As the laser fluence is increased in steps of 0.05 J/cm$^2$ from 0.45J/cm$^2$ to 0.55J/cm$^2$, the microbump collapses forming the micron-sized holes at the centre of laser irradiated spot. In fig. 4(a-d), it can also be observed that the cracks of the sputter deposited film surrounding the laser irradiated spot smoothen as a result of localized melting and vapor recoil pressure.

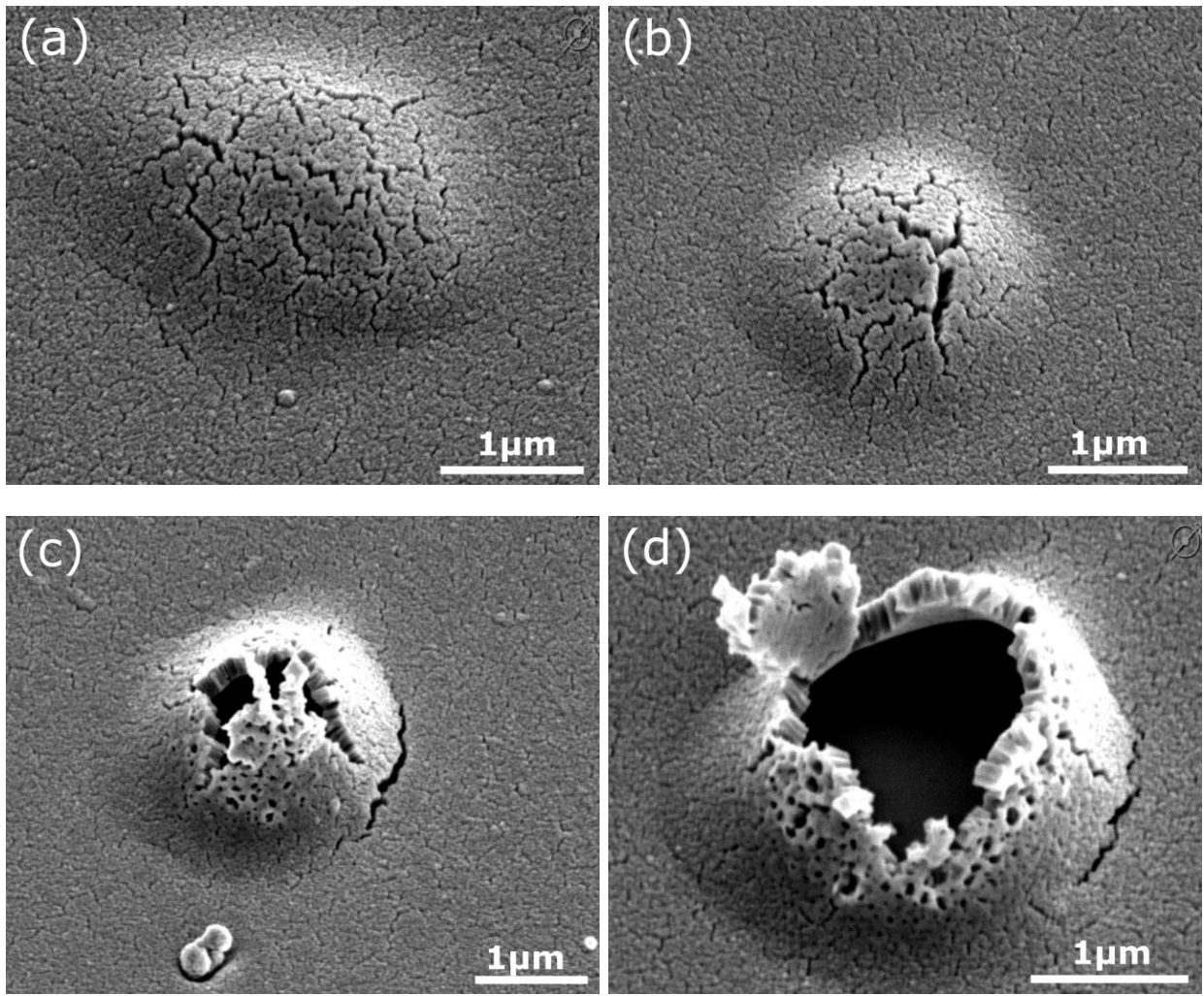

**Figure 5.** SEM images illustrating microbump formation on sputtered Ni films as a function of the energy density of the laser pulse for an irradiation spot size of 5μm: **(a)** 0.4 J/cm$^2$, **(b)** 0.45 J/cm$^2$, **(c)** 0.5 J/cm$^2$ and **(d)** 0.55 J/cm$^2$ single pulse.

The cracks and voids morphology of the sputtered films is not desirable, if a controllable formation of smooth microbump structures that are not fractured is to be obtained. Therefore, we focused our attention to methods for film deposition that result in smoother and compact Ni films. We were able to deposit much smoother Ni thin film of same thickness (200nm) by vacuum evaporation as can be seen from the next several sets of SEM images.

*3.3 Ni microbumps obtained on vacuum evaporated Ni thin film*

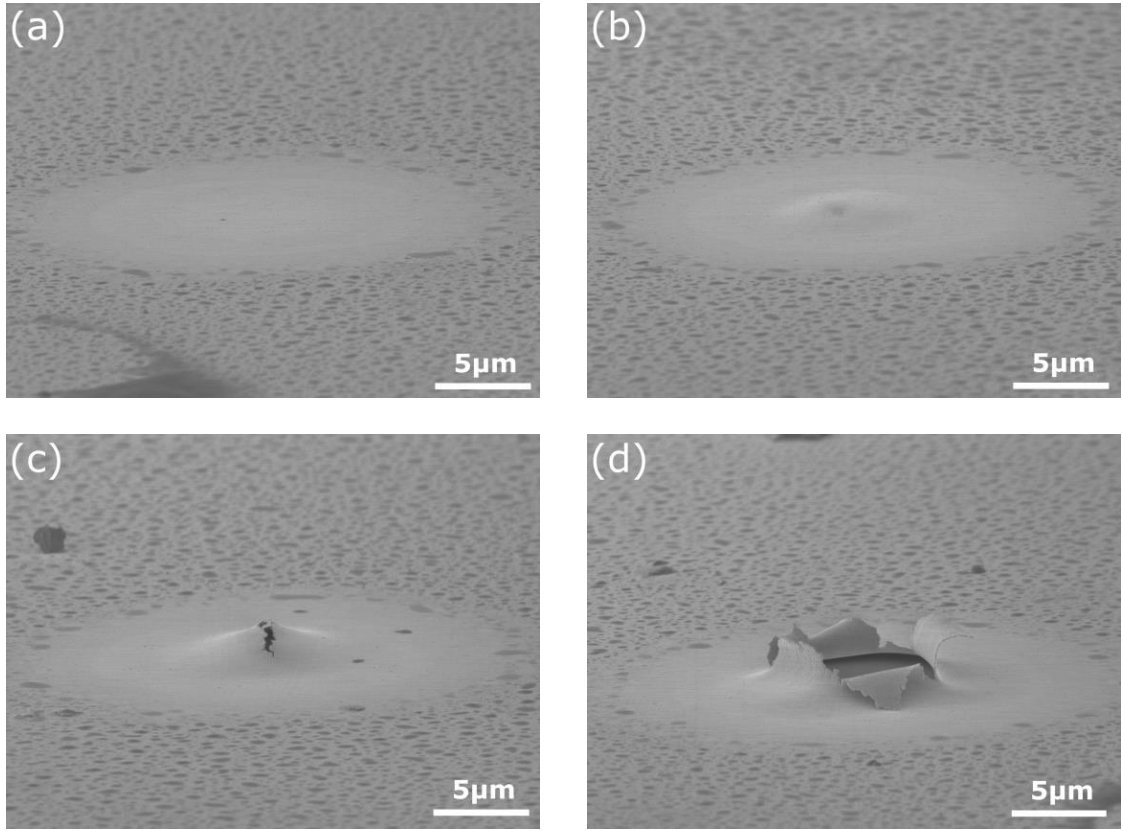

**Figure 6.** SEM images illustrating microbump formation on Ni films deposited by vacuum evaporation of thickness 100nm as a function of the energy density of the laser pulse for an irradiation spot size of 5µm: **(a)** 0.45 J/cm$^2$, **(b)** 0.5 J/cm$^2$, **(c)** 0.55 J/cm$^2$ and **(d)** 0.6 J/cm$^2$ single pulse.

Fig. 6 shows SEM images of Ni film of 100nm thickness deposited by vacuum evaporation. Porous grain boundaries can be observed for this thickness which suggests that a relatively thicker film deposition is required. Laser irradiation with fluence 0.45J/cm$^2$ causes the Ni material to thermally expand just enough causing these porous boundaries to merge, resulting in a smooth and uniform film at the laser spot. As the fluence is increased to 0.5J/cm$^2$ (Fig. 6 (b)), a small protrusion starts to form, eventually breaks at the tip of the protrusion due to lack of material beneath for expansion at a laser fluence of 0.55J/cm$^2$ (Fig. 6(c)). A further increase in fluence to 0.6J/cm$^2$ results in an ablated spot with the previously protruded section disintegrated into flakes.

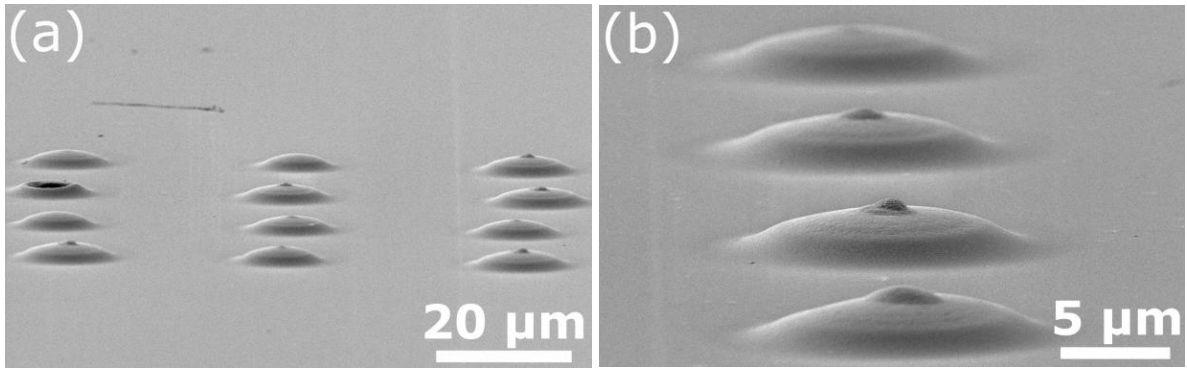

**Figure 7.** **(a)** SEM image of microbumps on evaporated Ni film, **(b)** a higher-magnification image of the microbumps is shown in the right panel.

Fig. 7 shows SEM images of microbumps formed as a result of single-pulse laser irradiation with 0.45 J/cm$^2$ on Ni film deposited by vacuum evaporation deposition. It also reveals the smooth nature of the film as compared to RF sputtering deposition. The grain boundaries are marginal (and impossible to see on this scale) as compared to the well-established boundaries on sputtered Ni film. In the case of these evaporated films, the lateral heat flow is more significant causing faster self-cooling and more substantial heating of the region immediately surrounding the irradiated spot. Perhaps even more importantly, the continuous nature of the film material (vs. voids and cracks in the sputtered film case) results in radially symmetric thermal expansion and deformation, the amount of which can be controlled (within certain limits) by the laser pulse energy. *Thus, reproducible microbumps with different sizes can be obtained, depending on the laser pulse energy.* The center of the spot is, again, the hottest and it can melt within the entire central part, followed by solidification that produces the central tip, which is somewhat similar to the case of gold films irradiation [21], [24], [25].

Fig. 8(a) shows a SEM scan of a 4x3 array of Ni microbumps obtained using a 75μm pin-hole mask resulting in a demagnified laser spot of 7.5μm. For the laser energy of 0.45 J/cm$^2$, a 90% yield is obtained owing to slight fluctuations in the laser energy. The microbump base diameter is about 7.5μm with a vertical height of 3μm.

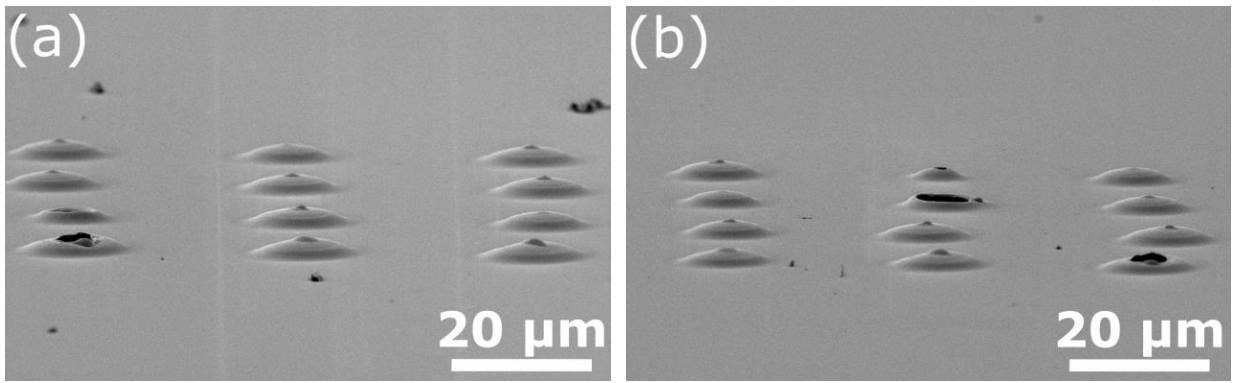

**Figure 8 (a).** SEM scan of a 4x3 array of microbumps on evaporated Ni film with laser energy of 0.45J/cm$^2$, **(b)** with laser energy of 0.5J/cm$^2$.

Fig. 8(b) is a SEM scan of a 4x3 array of Ni microbumps obtained for the laser energy of 0.5 J/cm$^2$, with a 83% yield of microbump formation. Since the melting temperature of Ni is around 1450$^0$C, the laser-generated heat does not fully melt the smooth nickel surface. Instead, a small portion of the surface enters molten state, rapidly solidifies and because of large difference in temperature gradient, the solidified bump breaks. This behavior is characteristic of metals which have a very high melting point. Hence, a 80-90% can be considered a reasonably good outcome of formation of microbumps on evaporated nickel film but this number can be improved by using more stable laser pulse energy.

For laser energies greater than 0.5 J/cm$^2$, ablation of the metal surface takes place as shown in Fig. 9. For a 4x3 array of laser spots with laser energy of 0.55 J/cm$^2$, more laser spots result in ablation, resulting in a 58% yield.

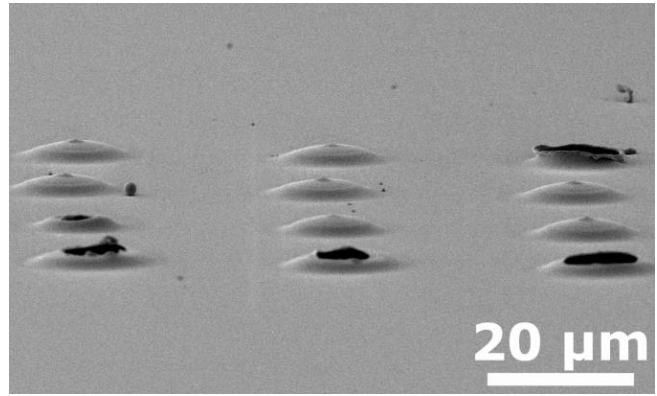

**Figure 9.** SEM scan of 4x3 array of microbumps on evaporated Ni film with laser energy of 0.55J/cm$^2$.

The model developed in Ref. [20] assumed microstructures that are hemispherical. It is clear from Fig. 10 that this corresponds to microbumps that were actually fabricated on evaporated nickel films, i.e., the assumption was reasonably accurate.

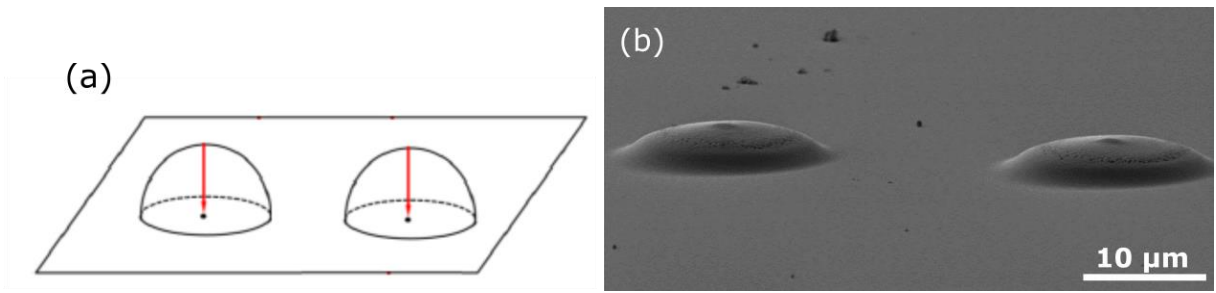

**Figure 10.** Comparison of **(a)** idealized hemispherical microstructures in ref.[20] with **(b)** fabricated microbumps on evaporated Ni film.

*3.4 Measurement of L and Q using VNA*

**Table 1.** Measured L and Q of the fabricated inductor stack

| Microstructuring | L | Q |
|---|---|---|
| NO | 0.42nH | 9.03 |
| YES (8 x 300 array) | 0.45nH | 10.15 |
| YES (12 x 300 array) | 0.46nH | 10.58 |

From table 1, it can be observed that microstructuring results in 7% increase in L and 9.5% increase in Q for a 8 x 300 array (12% of the total sample area) microstructuring. A slight increase in microstructuring (12 x 300 array) occupying 19% of the total sample area results in 9% increase in L and 10.5% increase in Q when compared to an inductor sample without microstructuring. These results indicate that a larger fraction of microstructured area (one can extrapolate up to 100% of the area) would provide even more significant increases in L and Q. Even more significant changes can be expected if the ferromagnetic metal film is changed from Ni to other, higher-permeability metals or alloys [20].

## 4. CONCLUSION

Conditions for the reproducible laser-based fabrication large arrays of microbumps on Ni films on glass substrates have been identified. The conditions and mechanism of formation appears to be similar to what was observed in earlier work on other metal films and deduced to be significantly dependent on the deposition techniques. On-chip spiral inductors with microstructured Ni film were fabricated and an improvement in the electrical parameters (L and Q) was observed. Compared to non-structured Ni film, a 9% increase in L and 10% increase in Q were measured using laser-microstructured Ni film. The nanosecond laser irradiation proves to be an efficient technique in enhancing the performance of on-chip spiral inductor, especially at high operating frequency, which is significant for GHz applications.

ACKNOWLEDGEMENT

The authors like to thank the department of Electrical Engineering and Computer Science (EECS) and the Center for Material Synthesis and Characterization (CMSC) at the University of Toledo for support to carry out this work. We would like to also thank Imaging Systems Technology, Toledo, Ohio, for help and support with the electrical characterization.